\documentclass[12pt,preprint]{aastex}

\slugcomment{ApJS}

\shorttitle{Photoionization and Recombination of Ni XXVI and Ni XXVII}
\shortauthors{Sultana N. Nahar}

\def \etal {et al.\thinspace}
\newcommand{\en}{\mbox{$n$}\ }
\newcommand{\el}{\mbox{$\ell$}\ }

\begin{document}


\title{Electron-Ion Recombination Rate Coefficients and Photoionization
Cross Sections for Astrophysically Abundant Elements. IX.
Ni XXVI and Ni XXVII for UV and X-ray modeling}


\author{Sultana N. Nahar\altaffilmark{} }
\affil{Department of Astronomy, The Ohio State University, Columbus,
    OH 43210}

\email{nahar@astronomy.ohio-state.edu}






\begin{abstract}
The inverse processes of photoionization and electron-ion recombination 
of (h$\nu$ + Ni XXVI $\leftrightarrow$ Ni XXVII + e) and (h$\nu$ + Ni 
XXVII $\leftrightarrow$ Ni XXVIII + e) are studied using the unified
method for the total recombination. The method subsumes both the
radiative and di-electronic recombination processes and enables
self-consistent sets of results for photoionization and electron-ion
recombination by using the same wavefunction for these inverse 
processes. Photoionization cross sections ($\sigma_{PI}$), recombination 
cross sections ($\sigma_{RC}$), recombination collision strengths 
($\Omega_{RC}$), and recombination rate coefficients ($\alpha_{RC}$) are 
obtained for ionization balance and spectral analysis of UV and X-ray 
lines. Level-specific photoionization cross sections and recombination 
rates are presented to enable accurate computation of 
recombination-cascade matrices for all fine structure levels 
\en (\el SLJ) up to \en $\leq 10$: 98 bound fine structure levels of 
Ni XXVI with 0 $\leq l\leq$ 9, 0 $\leq L \leq$ 11, 1/2 $\leq J \leq$ 
17/2, and 198 levels of Ni XXVII with 0 $\leq l\leq$ 9, 0 $\leq L \leq$ 
14, $0 \leq J \leq 10$. Total $\alpha_{RC}$ for Ni~XXVI and Ni~XXVII
are compared with the existing values with very good agreement. Total 
recombination rate coefficients for the hydrogen-like recombined ion, 
Ni XXVIII, are also presented. The calculations are carried out in 
relativistic Breit-Pauli R-matrix (BPRM) approximation with inclusion 
of radiation damping of resonances. With consideration of all details of 
the processes, the results, which include the level specific 
$\sigma_{PI}$ and $\alpha_{R}$ calculated for the first time, should be 
the most accurate for these ions. 
\end{abstract}


\section{INTRODUCTION}

From a study of photoionization and electron-ion recombination for
Li- and He-like ions, similar to the earlier ones for C~IV and C~V 
(Nahar \etal 2000), O VI, O VII (Nahar and Pradhan 2003), Fe XXV 
and Fe XXV (Nahar et al. 2001), results of cross sections and rates
are presented for Ni~XXVI and Ni~XXVII. These inverse atomic processes
of Li-like and He-like ions are of particular interest in the X-ray 
astronomy for analysis of new observations by space based observatories 
such as the Chandra X-ray Observatory and XMM-Newton, at photon 
energies and temperatures prevalent in high-temperature sources such as 
AGN, supernova remnants, hot stellar coronae etc. (e.g. Canizares \etal 
2000). 

The Li-like and He-like ions show distinctive features, compared to 
those of higher multi-electron ions, in both photoionization and 
electron-ion recombination in (i) displaying a featureless smooth 
structure until at high energies when narrow and dense resonances 
appear in separated $n$-complexes, and (ii) the low-n resonances are 
radiationally damped because of higher radiative decay rates. It is 
important to consider relativistic fine structure effect explicitly in 
the theoretical formulation to resolve the narrow resonances and include 
radiation damping of resonances. Photoionization cross sections, 
electron-ion recombination cross sections, and rate coefficients are 
obtained in the unified method for the total electron-ion recombination, 
accounting both radiative and di-electronic recombination (RR and DR)
processes, in the frame of relativistic Breit-Pauli R-matrix (BPRM) 
method (Nahar \& Pradhan 1992, 1994, Nahar 1996, Zhang et al. 1999). 
The unified method provides a a single set of recombination rate 
coefficents taking account of both RR and DR in an ab initio manner
for the entire temperature range of applications. 
The aim of the present series of reports (e.g. Nahar and 
Pradhan 1997) is at studying and presenting accurate atomic parameters 
for photoionization and total (e+ion) recombination for astrophysical 
models on a variety of applications.

\section{THEORY}

The unified method of electron-ion recombination (Nahar \& Pradhan
1992, 1994) yields self-consistent sets of atomic parameters for
the inverse processes of photoionization and recombination for atoms 
and ions. Photorecombination of an incident electron with the target 
ion may occur through (i) non-resonant, background continuum, or radiative 
recombination (RR),
\begin{equation}
e + X^{++} \leftrightarrow  h\nu + X^+,
\end{equation}
which is the inverse process of direct photoionization, or (ii) through
a two-step recombination process via autoionizing resonances, i.e.
dielectronic recombination (DR):
\begin{equation}
e + X^{++} \leftrightarrow (X^+)^{**}  \leftrightarrow  \left\{
\begin{array}{c} (i) \ e + X^{++} \\ (ii) \  h\nu + X^+ \end{array}
\right. ,
\end{equation}
The quasi-bound doubly-excited autoionizing state, $(X^+)^{**}$, leads 
either to (i) autoionization, a radiation-less transition to a lower 
target state with the electron going into a continuum, or (ii) radiative
stabilization to a recombined bound state via radiative decay of the 
ion core, usually to the ground state, with the capture of the electron. 

The unified method subsumes both the RR and DR processes. It considers 
photoionization from and recombination into the infinity of levels of 
the (e~+~ion) system. These recombined levels are divided into two groups:
group (A) bound levels with $n \leq n_o$ and  all possible fine structure 
$J\pi$ symmetries, and group (B) levels $ n_o < n \leq \infty $; where 
$n_o$ is the typically 10. 

Photoionization and recombination calculations are carried out in detail 
for all group (A) levels in close coupling (CC) approximation (e.g. 
Seaton 1997). In the CC approximation the target ion (core) is
represented by an $N$-electron system. The total wavefunction, $\Psi(E)$,
of the ($N$+1) electron-ion system of symmetry $J\pi$ is
represented in terms of an expansion of target eigenfunctions as:
\begin{equation}
\Psi(ion+e;E) = A \sum_{i} \chi_{i}(ion)\theta_{i} + \sum_{j} c_{j} \Phi_{j}
(ion+e),
\end{equation}
where $\chi_{i}$ is the target wavefunction for a specific level
$J_i\pi_i$ and $\theta_{i}$ is the wavefunction for the ($N$+1)-th 
electron in a channel labeled as $S_iL_i(J_i)\pi_ik_{i}^{2}\ell_i(\ J\pi)$, 
$k_{i}^{2}$ being its incident kinetic energy. The $\Phi_j$'s are the 
correlation functions of the ($N$+1)-electron system that account for 
short range correlation and the orthogonality between the continuum and 
the bound orbitals.

In relativistic BPRM calculations, developed under the Iron Project 
(IP, Hummer et al. 1993), the set of ${SL\pi}$ are recoupled
for $J\pi$ levels of (e + ion)-system, followed by diagonalisation of
the Hamiltonian, $H^{BP}_{N+1}\mit\Psi = E\mit\Psi$,
where the BP Hamiltonian is
\begin{equation}
H_{N+1}^{\rm BP}=H_{N+1}^{NR}+H_{N+1}^{\rm mass} + H_{N+1}^{\rm Dar}
+ H_{N+1}^{\rm so}.
\end{equation}
The first term, $H_{N+1}^{NR}$, is the nonrelativistic Hamiltonian,
\begin{equation}
H_{N+1}^{NR} = \sum_{i=1}\sp{N+1}\left\{-\nabla_i\sp 2 - \frac{2Z}{r_i}
        + \sum_{j>i}\sp{N+1} \frac{2}{r_{ij}}\right\},
\end{equation}
and the additional one-body terms are 
\begin{equation}
\begin{array}{l}
the~mass~correction~term,~H^{\rm mass} = -{\alpha^2\over 4}\sum_i{p_i^4},\\
the~Darwin~term,~H^{\rm Dar} = {Z\alpha^2 \over 4}\sum_i{\nabla^2({1
\over r_i})}~ and \\
the~spin-orbit~interaction~term,~H^{\rm so}= Z\alpha^2 \sum_i{1\over r_i^3}
{\bf l_i.s_i},
\end{array}
\end{equation}
respectively. The spin-orbit term splits the $LS$ term into fine structure 
components.

The positive and negative energy states (Eq. 4) define continuum or
bound (e~+~ion) states. $E = k^2 >$ 0 for continuum (scattering) channels 
and $E <$ 0 for bound states. The reduced matrix element for the 
bound-free transition,
\begin{equation}
<\Psi_B || {\bf D} || \Psi_{F}>,
\end{equation}
can be obtained from the continuum wavefunction ($\Psi_{F}$) and the 
bound wavefunction ($\Psi_B$). {\bf D} is the dipole operator, 
${\bf D}_L = \sum_i{\bf r_i}$, in length form, the sum is on  the number 
of electrons. The photoionization cross section is obtained as
\begin{equation}
\sigma_{PI} = {1\over g_i}{4\pi^2\over 3c}\omega{\bf S},
\end{equation}
where $g_i$ is the statistical weight factor of the initial bound state 
and ${\bf S}~=~|<\Psi_B || {\bf D} || \Psi_F >|^2$ is the dipole line 
strength, For highly charged H- and the He-like recombining ions, the 
probability of radiative decay of 
an autoionizing state is often comparable to that of autoionization 
(typically 10$^{12}$ - 10$^{14}~sec^{-1}$, as discussed in Nahar \etal 
2000). With strong dipole allowed $ 2p \longrightarrow 1s$ and 
$ 1s2p \ (^1P^o_1) \longrightarrow 1s^2 \ (^1S_0)$ transitions (e.g. 
Table 2) autoionizing resonances are radiatively damped to a significant 
extent. The radiative damping effect of all near-threshold resonances, 
up to effective quantum number $\nu \leq 10$, are considered using a
resonance fitting procedure (Sakimoto et al. 1990, Pradhan and Zhang 1997, 
Zhang \etal 1999).

The photo-recombination cross section, $\sigma_{\rm RC}$, is related 
to photoionization cross section, $\sigma_{\rm PI}$, through principle 
of detailed balance (Milne relation) as
\begin{equation}
\sigma_{\rm RC}(\epsilon) =
{\alpha^2 \over 4} {g_i\over g_j}{(\epsilon + I)^2\over \epsilon}
\sigma_{\rm PI}.
\end{equation}
\noindent
$\alpha$ is the fine structure constant, $\epsilon$ 
is the photoelectron energy, $g_j$ is the statistical weight factor of
the recombined ion and $I$ is the ionization potential. $\sigma_{\rm RC}$ 
are computed from the photoionization cross sections 
at a sufficiently large number of energies to delineate the non-resonant
background and the autoionizing resonances, thereby representing both
radiative and the dielectronic recombination (RR and DR) processes.
In the unified treatment the photoionization cross sections,
$\sigma_{\rm PI}$, of a large number of bound states (group A) -- all
possible states with $n \leq n_{\rm o}\sim 10$ -- are obtained as
described above. It is assumed that the recombining ion is in the ground
state, and recombination can take place into the ground or any of the
excited recombined (e+ion) states.

Recombination rate coefficients of individual recombined levels are 
obtained by convolving recombination cross sections over Maxwellian 
electron distribution $f(v)$ at a given temperature as,
\begin{equation}
\alpha_{RC}(T) = \int_0^{\infty}{vf(v)\sigma_{RC}dv}.
\end{equation}
Contributions from the low-\en group (A) bound levels are added for the 
total recombination rate coefficient, $\alpha_{RC}$ and for the total
recombination cross sections, $\sigma_{\rm RC}$. 

Group (B) levels, $n_{\rm o} < n \leq \infty$, are treated through 
quantum defect theory of DR within close the coupling approximation 
(Nahar \& Pradhan 1992, 1994). A generally valid approximation made 
in recombination to group (B) levels is that the background 
contribution is negligible, and DR is the dominant process in the 
region below the threshold of convergenece for high-\en resonances. 
To each excited threshold of the core, $J_i\pi_i$, belongs an 
infinite series of ($N$+1)-electron levels, $J_i\pi_i\nu\ell$, to 
which recombination can occur. For the high $\nu$ levels DR dominates 
while the background RR is small and, hence DR dominates the total 
recombination. The contributions from these levels are added by 
calculating the DR collision strengths, $\Omega_{\rm DR}$, an 
extension (Nahar and Pradhan 1994) of the precise theory of 
radiation damping by Bell and Seaton (1985):
\begin{equation}
\Omega(DR) = \sum_{SL\pi} \sum_n (1/2)(2S+1)(2L+1) P^{SL\pi}_{n}.
\end{equation}
$P^{SL\pi}_n$ is the DR probability in entrance channel $n$, obtained
as $P_{n}^{SL\pi}({\rm DR}) = \mbox{\boldmath $(1 - {\cal S}_{ee}^{\dagger}
{\cal S}_{ee})_{n}$}$, where ${\bf \cal S}_{\rm ee}$ is the matrix for 
electron scattering {\em including} 
radiation damping. The recombination cross section, $\sigma_{RC}$ in
Megabarns (Mb), is related to the collision strength, $\Omega_{\rm RC}$, as
\begin{equation}
\sigma_{RC}(i\rightarrow j)({Mb}) = \pi \Omega_{RC}(i,j)/(g_ik_i^2)
(a_o^2/1.\times 10^{-18}) ,
\end{equation}
where $k_i^2$ is the incident electron energy in Rydbergs. Since
$\sigma_{RC}$ diverges at zero electron energy, the total collision 
strength, $\Omega$, is used in the recombination rate calculations.

The group (A) levels in the unified method may not be necessarily 
restricted up to \en = 10. It can readily be extended to higher n. However, 
detailed calculations for photoionization cross sections of high-\en 
($>$ 10) levels is unnecessary since they approach hydrogenic behavior. 
Background photoionization cross sections for these high-\en group (B) 
levels are computed hydrogenically to contribute to the total
 recombination rate. This high-n background contribtution is referred 
to as the ``high-\en top-up" contribution (Nahar 1996).

\section{COMPUTATIONS}

The calculations for photoionization and electron-ion recombination span 
several stages of computation starting with obtaining the target 
(core ion) wavefunction through configuration interaction atomic structure 
calculations. The wavefunction expansion for Ni XXVI consists of 17 fine 
structure levels of configurations $1s^2$, $1s2s$, $1s2p$, $1s3s$, 
$1s3p$, and $1s3d$ of target Ni XXVII. The levels, along with their 
relative energies, are given in Table 1. The set of correlation 
configurations in the atomic structure calculations are also given in 
Table 1. The orbital wavefunctions are obtained from the atomic 
structure code SUPERSTRUCTURE (Eissner et al. 1974). The Thomas-Fermi 
scaling paramenter ($\lambda_{nl}$) for each orbital is taken to be 1. 
The second term of the wavefunction in Eq. (3), which contains bound 
state correlation functions for Ni XXVI, includes all possible 
($N$\,+\,1)-particle configurations from 0 to maximum orbital 
occupancies as 2s$^2$, 2p$^2$, 3s$^2$, 3p$^2$, 3d$^2$, $4s$ and $4p$. 
The energies in Table 1 are observed values 
(from the NIST website: www.nist.gov). Radial integrals for the 
partial wave expansion in Eq.\,3 are specified for orbitals 
$0\leq\ell\leq 9$, with a R-matrix basis set of 40 `continuum' functions 
for Ni XXVI. Computations are carried out for all angular 
momenta, 0 $\leq L\leq$ 11, 1/2 $\leq J\leq$ 17/2 for Ni XXVI.

The wavefunction expansion of Ni XXVII consists of 16 fine structure
levels of configurations. $1s$, $2s$, $2p$, $3s$, $3p$, $3d$, $4s$, $4p$, 
$4d$ and $4f$ of Ni XXVIII, as given in Table 1. There are no correlation 
configurations, and the scaling parameter for each orbital is taken to 
be unity. The orbital wavefunctions and level energies are obtained from 
SUPERSTRUCTURE. The energies in Table 1 are the calculated ones. The 
bound state correlation functions, in the second term of the wavefunction,
include all configurations from 0 to maximum orbital occupancies:
$1s^2$, 2s$^2$, 2p$^2$, 3s$^2$, 3p$^2$, 3d$^2$, $4s^2$ and $4p^2$,
$4d$ and $4f$. Radial integrals for the partial wave expansion are 
specified for orbitals $0\leq\ell\leq 9$, with a R-matrix basis set
of 30 `continuum' functions for Ni~XXVII. Computations are 
carried out for all angular momenta, 0 $\leq L\leq$ 14, 0 $\leq J\leq$ 
10 for Ni XXVII.

Relativistic BPRM calculations in intermediate coupling are carried
out in the close coupling approximation using the $R$-matrix package of
codes (Berrington et al. 1995). These are extensions of the Opacity 
Project codes (Berrington et al. 1987) to include relativistic effects 
(Scott \& Burke 1980, Scott \& Taylor 1982, Berrington et al. 1995), 
implemented under the Iron Project (Hummer \etal 1993). The energy levels 
were identified using code PRCBPID (Nahar and Pradhan 2000).

Both the {\it partial} and the {\it total} photoionization cross setions 
are obtained for all bound levels. Coupled channel calculations for
$\sigma_{\rm PI}$ include both the background and the resonance
structures (due to the doubly excited autoionizing states) in the cross
sections. Radiation damping of resonances up to $n = 10$ are included
through use of the extended codes STGF and STGBF (Nahar \& Pradhan 1994,
Zhang \etal 1999). The BPRM calculations are carried out for each total
angular momentum symmetry $J\pi$, corresponding to a set of fine
structure target levels $J_t$.  Radiation damping of resonances
within the close coupling BPRM calculations are described in Zhang \etal
(1999, and references therein). The program PBPRAD is used to extend the
{\it total} photoionization cross sections in the high energy region,
beyond the highest target threshold in the close coupling wavefunction
expansion of the ion, by a `tail' using a fitting formula as well as
Kramers formula $\sigma(E) =\sigma_{PI}^o(E^o/E)^3$ where $\sigma^o$ is 
the last tabulated cross section at energy $E^o$ above all target 
thresholds, as described in Nahar and Pradhan (1994).

In the higher energy region, $\nu_o < \nu \leq \infty$, below each
target threshold where the resonances are narrow and dense and
the background is negligible, we compute detailed and resonance
averaged DR cross sections (Bell and Seaton 1985, Nahar and Pradhan 1994). 
The radiative decay probabilities used for the ions are given in 
Table 2. The DR collision strengths in BPRM approximation are obtained 
using extensions 
of the $R$-matrix asymptotic region code STGF (Nahar \& Pradhan 1994, 
Zhang \etal 1999). It is necessary to use a very fine energy mesh in 
order to delineate the resonance structures.

Level-specific recombination cross sections, $\sigma_{\rm RC}(i)$, into
various bound levels $i \equiv$ \en(SLJ) of the recombined (e~+~ion) 
system, are obtained from {\it partial} photoionization cross sections 
$ \sigma_{\rm PI}(i,g)$ of the level $i$ into the ground level $g$ of 
the recombining ion. These detailed photo-recombination cross sections 
are calculated in the energy region from the threshold energy up to 
$E(\nu = \nu_o \approx 10.0)$, where $\nu$ is the effective quantum 
number relative to the target level of the recombining ion. The 
resonances up to $\nu \leq \nu_o$ are delineated with a fine energy
mesh. The electrons in this energy range generally recombine to a large
number of final (e~+~ion) levels. The level specific rates are 
obtained for energies going up to infinity. Recombination cross sections 
are computed for all coupled symmetries and levels, and summed to obtain 
the total $\sigma_{\rm RC}$ using the program, RECXS (Nahar \etal 2000).  

The program RECXS sums up the level specific rates for the total
recombination rates, which also includes the contributions from the 
resonant high-n DR of $\nu_o < \nu < \infty$. As an additional check 
on the numerical calculations, the total recombination rate coefficients, 
$\alpha_R$, are also calculated from the total recombination collision 
strength, $\Omega_{RC}$, obtained from all the photoionization cross 
sections, and the high-n DR collision strengths. The agreement between 
the two numerical approaches is within a few percent.

The small background (non-resonant) contribution from the high-\en 
states ($10 < n \leq \infty$) to total recombination is included as 
the "top-up" part, computed in the hydrogenic approximation (Nahar 
1996). This contribution is important at low temperatures as the
recombination rate is dominated by the RR, but negligible at high 
temperatures.

\section{RESULTS AND DISCUSSION}

The inverse processes of photoionization and recombination of (Ni XXVI 
+ $h\nu$ $\leftrightarrow$ Ni XXVII + e), and (Ni XXVII + $h\nu$ 
$\leftrightarrow$ Ni XXVIII + e) are studied in detail. Total 
recombination rate coefficients for the hydrogenic Ni XXVIII are 
also presented alongwith those of Ni XXVI and Ni XXVII in Table 3 for 
completeness. 

Both the total and the partial photoionization cross sections,
including the autoionizing resonances, are obtained for Ni~XXVI and 
Ni~XXVII for the first time and are available electronically.
The {\it total} photoionization cross sections correspond to leaving 
the core ion in various excited states and are needed in astrophysical 
applications, such as in ionization balance calculations. The 
{\it parital} cross sections for leaving the core in the ground level
are needed for applications such as for recombination rate coefficients. 

The total unified recombination collision strengths ($\Omega_{RC}$), 
cross sections ($\sigma_{RC}$), and recombination rate ($\alpha_{RC}(E)$) 
with electron energies are presented for 
Ni~XXVI and Ni~XXVII, and data are available electronically. Total 
unified recombination rate coefficient, $\alpha_{RC}(T)$, are computed 
in two different ways to enable numerical checks: (i) from
the sum of the level-specific rate coefficients and the high-n DR
contribution, and (ii) from total collision strengths calculated from
photoionization cross sections directly, and the DR contribution. The
differences between the two are typically within a few percent, thus
providing a numerical and self-consistency check particularly on the
resolution of resonances.  

Level-specific and total recombination rate coefficients, $\alpha_i$ 
(\en SLJ, \en $\leq$ 10) and $\alpha_{RC}(T)$ respectively, are obtained 
using the BPRM unified treatment for Ni~XXVI and Ni~XXVII. 
Calculations of the recombination-cascade contributions for 
important lines require accurate atomic parameters for fine structure 
levels up to fairly high \en levels, as reported here.
Level-specific recombination rate coefficients of all bound levels 
and the summed total recombination rate coefficients ($\alpha_R(T)$) 
for recombined into infinite number of bound states are presented. 
The level-specific rate coefficients are obtained for the first time.
Existing data are available only for individual total RR and DR rates.

Important features in photoionization and electron-ion recombination 
for each ion are discussed separately below.

\subsection{Ni XXVI}

A total of 98 bound levels are found for Ni XXVI with n $\leq$ 10, 
0 $\leq l\leq$ 9, 0 $\leq L \leq$ 11, and total angular momentum 
of $1/2 \leq J \leq 17/2$ (Nahar 2002).

\subsubsection{Photoionization}

Cross sections ($\sigma_{PI}$) for both {\it total} and {\it partial}
photoionization are obtained for all 98 bound levels of Ni XXVI. 

Figs. 1(a) and (b) present the ground state photoionization cross section 
for Ni XXVI ($1s^2 \ 2s \ ^2S_{1/2}$). The top panel (a) presents
the {\it total} photoionization cross section summed over the various
target thresholds for ionization, and the bottom panel (b) presents
the {\it partial} cross sections of the ground level into the ground
$1s^2(^1S_0)$ level of residual ion Ni XXVII. The resonances at high
energies belong to the Rydberg series converging on to the $n=2,3$ 
levels of core Ni~XXVII. These are the well known KLL, KLN resonances, 
for example, as discussed by Nahar \etal (2003). Since 
the first excited levels of n=2 thresholds of the core ion Ni~XXVII 
lie at high energies, the cross sections decrease monotonically over 
a large energy range before the Rydberg series of resonances appear. 
The {\it total} and the {\it partial} cross sections are identical 
below the first excited level of the residual ion beyond which total 
$\sigma_{PI}$ increases due to added contributions from excited 
channels (Fig. 1). The distinct difference between the total and partial 
cross sections Fig. 1 comes from the contribution
of channels with excited n=2 thresholds. The K-shell ionization
jump at the n = 2 target levels in total $\sigma_{PI}$ is due to
inner-shell photoionization:
$$ h\nu + Ni~XXVI (1s^2 \ 2s) \longrightarrow e \ + Ni~XXVII (1s2s ,1s2p) \ .$$
In X-ray photoionization models inner-shell edges play an important
role in overall ionization rates.

Fig. 2 presents partial photoionization cross sections of excited
Rydberg series of levels $1s^2np(^2P^o_{1/2})$, with 2 $\leq n \leq$ 7, 
of Ni~XXVI. The figure illustrates the resonant structures at higher 
energies. especially the photoexcitation-of-core (PEC) resonances at 
energies associated with dipole transitions in the core ion. PEC 
resonances in photoionization cross sections are seen in all excited 
bound levels of Ni~XXVI, as in Fig. 2, at photon energies 570.79, 573.67,
674.15, and 674.98 Ry due to core excitations to levels $1s2p(^3P^o_1)$, 
$1s2p(^1P^o_1)$, $1s3p(^3P^o_1)$, and $1s3p(^1P^o_1)$ of Ni~XXVII. 
At these energies the core ion goes through an allowed transition, 
while the outer electron remains as a `spectator' in a doubly-excited 
resonance state, followed by autoionization into the ground level of 
the core. The effect is more prominent for cross sections of higher 
excited levels. These resonances depict the non-hydrogenic behavior 
of cross sections of excited levels and contribute to features in 
level-specific recombination rates.

\subsubsection{Electron-ion Recombination}

The collision strength for electron-ion recombination, $\Omega_{RC}$, 
is related to recombination cross sections, $\sigma_{RC}$ (Eq. (10)), 
and show similar features. However, $\sigma_{RC}$ blows up at zero 
electron energy. Hence the total unified photorecombination collision 
strength, $\Omega_{RC}$, for Ni XXVI is presented in Fig. 3(a). 
$\Omega_{RC}$, similar to photoionization cross sections $\sigma_{PI}$, 
decays smoothly with energy, before the emergence of resonance 
complexes at high energies. However, $\Omega_{RC}$ is more complex than 
single level $\sigma_{PI}$ as it is obtained from summed contributions of 
$\sigma_{PI}$ of all levels. The resonance complexes in $\Omega_{RC}$ 
(marked in the figure) are KLL, KLM, KLN etc. going up to the n = 2 threshold 
and KMM, KMN etc. going up to the n = 3 threshold, where KLL means 1s2l2l, 
KLM means 1s2l3l' etc. These resonances manifest themselves as 
di-electronic satellite lines observed in tokamaks, Electron-Beam-Ion-Traps 
(EBIT), ion storage rings and astrophysical sources. The KLL complexes 
have been well studied in previous works (e.g. Gabriel 1972, Pradhan and 
Zhang 1997, Zhang et al.  1999, Oelgoetz and Pradhan 2001), for various 
ions. 

Photorecombination rates in terms of photoelectron energy,
\begin{equation}
\alpha_{RC}(E) = v\sigma_{RC}(E),
\end{equation}
where $v$ is the photoelectron energy, an experimentally measurable 
quantity. Fig. 3(b) presents the expanded part of the resonances up to 
n = 2 threshold in $\alpha_{RC}(E)$ of Ni~XXVI. The observed shape 
corresponds to the detailed $\alpha_{RC}(E)$ convolved by the monochromatic
bandwidth of the experiment (e.g. for C III by Pradhan et al. 2001). 

Level-specific recombination rate coefficients $\alpha_R(T)$ of 98 
levels are presented for Ni XXVI. They correspond to all associated 
$J\pi$ levels $i \equiv n(SLJ)$ with n $\leq$ 10 and $\ell \leq 9$.
Fig. 4 presents $\alpha_i(T)$ into the eight lowest excited \en = 2 and 
3 levels of \en(SLJ): $2s~^2S_0$, $2p~^2P^o_{1/2,3/2}$,
$3s~^2S_0$, $3p~^2P^o_{1/2,3/2}$, and $3d~^2D_{3/2,5/2}$. These rates
are relatively smooth except for a small and diffuse DR `bump' at high
temperature.

Total recombination rates are given in Table 3. The main features 
are illustrated and compared with previously available data for
RR and DR rates in Fig. 5. The solid curve in the figure is the BPRM 
total unified $\alpha_R(T)$ and shows typical features. Starting with
a higher rate coefficient at very low temperature, due to the 
dominance of RR into an infinity of high-\en levels. $\alpha_R(T)$ 
decreases with increasing T until at high temperatures where it rises 
due to the dominance of DR, followed by a monotonc deccay.  

The present total unified recombination rate coefficients, $\alpha_R(T)$,
for Ni~XXVI are compared with RR rate coefficients (dash) by Verner and
Ferland (1996), DR rates (dotted) by Jacobs et al (1980) and DR rates
(chain dash) by Romanik (1988). Present rates agree very well with 
the previously calculated rates, especially with the sum of RR+DR 
(dot-dash) of Verner and Ferland and of Jacobs et al.

\subsection{Ni XXVII}

A total of 198 bound levels are found for Ni XXVII with n $\leq$ 10, 
0 $\leq l\leq$ 9, 0 $\leq L \leq$ 14, and total angular momentum 
of $0 \leq J \leq 10$.

\subsubsection{Photoionization}

Partial and total photoionization cross sections are presented for 
all 198 bound levels (n $\leq$ 10) of Ni~XXVII. Illustrative results 
are presented in Fig. 6 and 7.

Fig. 6 presents samples of level-specific photoionization cross section 
of Ni~XXVII. The ground $1s^2 \ (^1S_0)$ level cross sections are in the
topmost panel while the lower four panels present $\sigma_{PI}$ of the 
four lowest n = 2 excited levels of Ni~XXVII. These excited levels 
correspond to the prominent lines of K$\alpha$ complex in the X-ray 
emission of He-like ions; resonance (w: $1s^2 \ (^1S_0) \leftarrow 
1s2p (^1P^o_1)$), intercombination (y : $1s^2 \ (^1S_0) \leftarrow 
1s2p (^3P^o_{1})$), and forbidden (x: $1s^2 \ (^1S_0) \leftarrow 
1s2p (^3P^o_{2})$, and z: $1s^2 \ (^1S_0) \leftarrow 1s2s (^3S_1)$) 
lines. respectively, and yield valuable spectral diagnostics
of temperature, density, ionization balance, and abundances in the plasma
source.

In the photoionization cross section of the ground level $(1s^2 \ ^1S_0)$
of Ni~XXVII, the Rydberg series of resonances, KL and K\en ($n > 2$), 
begin at fairly high energies owing to the high $n=2$ excitation 
thresholds of Ni~XXVIII. However, the ground level $\sigma_{PI}$ of 
Ni~XXVII does not show a significant K-shell jump at $n=2$ threshold, as 
seen in Ni~XXVI. Nonetheless the K-shell ionization jump at the n = 2 
target levels 
$$ h\nu + Ni~XXVII (1s2s,1s2p) \longrightarrow e \ + Ni~XXVIII (2s \ ,2p) \ $$ 
can be seen clearly in the photoionization cross sections in Figs. 6(b)-(e).

Fig. 7 presents partial photoionization cross sections of the Rydberg 
series of J = 0 levels, $1sns(^1S_{0})$ where 2 $\leq n \leq$ 7, of 
Ni~XXVII ionizing into the ground level $1s(^2S_{1/2})$ of the core. 
The figure displays the PEC resonances at about 594, 704, 743 Ry
(marked by arrows in the top panel) due to core transitions to allowed
levels. The energy positions correspond to core levels 
$^2P^o_{1/2},^2P^o_{3/2}$ of configurations $2p$, $3P$, and $4p$.

\subsubsection{Electron-ion recombination}

The total unified photorecombination collision strength $\Omega_{RC}$ 
for Ni XXVII is presented in Fig. 8(a). It corresponds to summed 
contributions from photoionization cross sections of all levels and 
high-n DR. $\Omega_{RC}$ decays smoothly with energy until resonance 
complexes appear at very high energy, as expected for a He-like ion. 
The resonance complexes are LL, LM, LN etc. going up to the n = 2 threshold 
and MM, MN etc. going up to the n = 3 threshold, and NN, NO etc. going up 
to the n = 4 threshold; LL means 2l2l, LM means 2l3l' etc. The $n = 4$ 
resonances are too weak for any significant contributions. 

Fig. 8(b) presents an expanded part of the resonances up to the n = 2 
threshold of photorecombination rates ($\alpha_{RC}(E)$) varying with 
photoelectron energy. As mentioned before, $\alpha_{RC}(E)$ is 
a measurable quantity and the observed shape is usually convolved by 
the monochromatic bandwidth of the experiment.

Level-specific recombination rate coefficients are obtained for 198
levels of \en(SLJ) with 0  $ \leq J \leq$ 10 and \en $\leq$ 10.
Fig. 9 presents level specific rates for the ground and the $n = 2$
levels corresponding to the X-ray w, x, y, and z lines of Ni~XXVII. The
rates show a relatively smooth decrease with temperature.

The total unified recombination rate coefficients are given in Table 3. 
The main features are illustrated and compared with previously available 
data for RR and DR rates in Fig. 10. The BPRM total unified
$\alpha_R(T)$ (solid curve in the figure) shows typical features. The 
high recombination rate coefficients at very low temperature, due to 
the dominance of RR into an infinity of high-\en levels. decreases 
with increasing T until at high temperatures where it shows a "shoulder" 
due to the dominance of DR, which is followed by a monotonic decay. 

The present total unified recombination rate coefficients, $\alpha_R(T)$,
for Ni~XXVI are compared with RR rate coefficients (dash) by Verner and
Ferland (1996) and DR rates (dotted) by Jacobs et al (1980). Present 
rates agree very well with the previously calculated rates, especially 
with the sum of RR+DR (dot-dash) of Verner and Ferland and of Jacobs et al.

\section{CONCLUSION}

Extensive results from relativistic calculations for total and
level-specific photoionization and recombination cross sections and rates
are presented for Ni~XXVI and Ni~XXVII. These are of general interest in 
UV and X-ray spectroscopy of laboratory and astrophysical sources. 

The present level-specific data can be used to construct 
recombination-cascade matrices for Ni~XXVI and Ni~XXVII to obtain 
effective recombination rates into specific fine structure levels 
\en(SLJ) with $n \leq 10$ and $\ell \leq n-1 $ (e.g. Pradhan 1985). 
Present total unified recombination rates agree very well with the sum 
of RR and DR of previous calculations. It is expected for the He- and 
Li-like ions for which electron-electron correlation is weak resulting
in a small interference between RR and DR. However, the unified 
method for recombination provides level-specific rates of hundreds of 
levels and corresponding self-consistent photoionization cross sections 
of many bound levels, not obtainable by other existing methods. The 
present data is more than sufficient for extrapolation to high-n,$\ell$ 
necessary to account for all cascade contributions.

The available data includes:
(i) Photoionization cross sections, both total and partial, for bound fine
structure levels of Ni~XXVI and Ni~XXVII up to the $n = 10$ levels.
(ii) Total unified recombination rate coefficients for Ni~XXVI and Ni~XXVII, 
and level-specific recombination rate coefficients for levels up to 
$n = 10$, (iii) total unified recombination collision strength, cross
sections and rates with photoelectron energies of Ni~XXVI and Ni~XXVII.
Further calculations for other He-like and Li-like ions are in progress.
All photoionization and recombination data are available electronically
at: nahar@astronomy.ohio-state.edu.

%

\acknowledgments

This work was supported partially by NSF and NASA. The
computational work was carried out on the Cray SV1 at the Ohio Supercomputer
Center in Columbus, Ohio.

\clearpage

\begin{table}
\caption{Ion core levels in the eigenfunction expansions of Ni XXVI and 
Ni XXVII.}
\scriptsize
\begin{tabular}{rllrll}
\hline
&\multicolumn{2}{c}{Ni XXVII} & \multicolumn{3}{c}{Ni XXVIII} \\
& level & $E_t(Ry)$ & & level & $E_t(Ry)$ \\
\hline
1&1s$^2(^1{\rm S}_0)$     & 0.0    &
1&1s$(^2{\rm S}_{1/2})$   &   0.00 \\
2&1s2s$(^3{\rm S}_1)$     &568.2566   &
2&2p$(^2{\rm P}^o_{1/2})$ &593.654    \\
3&1s2p$(^3{\rm P^o}_0)$   &570.5865    &
3&2s$(^2{\rm S}_{1/2})$   &593.686   \\
4&1s2s$(^1{\rm S}_0)$     &570.792     &
4&2p$(^2{\rm P}^o_{3/2})$ &595.723   \\
5&1s2p$(^3{\rm P}^o_1)$   & 570.7938   &
5&3p$(^2{\rm P}^o_{1/2})$ &704.226    \\
6&1s2p$(^3{\rm P}^o_2)$   &572.2873  &
6&3s$(^2{\rm S}_{1/2})$   &704.238   \\
7&1s2p$(^1{\rm P}^o_1)$   &573.6669    &
7&3d$(^2{\rm D}_{3/2})$   &704.821   \\
8&1s3s$(^3{\rm S}_1)$     &673.4568    &
8&3p$(^2{\rm P}^o_{3/2})$ &704.823   \\
9&1s3p$(^3{\rm P}_0)$     &674.1039    &
9&3d$(^2{\rm D}_{5/2})$   &705.024   \\
10&1s3s$(^1{\rm S}_0)$    &674.1223   &
10&4p$(^2{\rm P}^o_{1/2})$&742.836    \\
11&1s3p$(^3{\rm P}^o_1)$  &674.1493  &
11&4s$(^2{\rm S}_{1/2})$  &742.841   \\
12&1s3p$(^3{\rm P}^o_2)$  &674.5945    &
12&4d$(^2{\rm D}_{3/2})$  &743.065   \\
13&1s3p$(^1{\rm P}^o_1)$  &674.9808  &
13&4p$(^2{\rm P}^o_{3/2})$&743.078  \\
14&1s3d$(^3{\rm D}_1)$    &676.624     &
14&4d$(^2{\rm D}_{5/2})$  &743.148   \\
15&1s3d$(^3{\rm D}_2)$    &676.632     &
15&4f$(^2{\rm F}^o_{5/2})$&743.162   \\
16&1s3d$(^1{\rm D}_3)$    &676.806     &
16&4f$(^2{\rm F}^o_{7/2})$&743.205   \\
17&1s3d$(^1{\rm D}_2)$    &676.819     &
 & \\
\hline
\multicolumn{6}{l}{Ni XXVII: Correlations -  $2s^2$, $2p^2$, $3s^2$,
$3p^2$, $3d^2$, $2s2p$,}\\
\multicolumn{6}{l}{$2s3s$, $2s3p$,$2s3d$, $2s4s$, $2s4p$, $2p3s$,
$2p3p$, $2p3d$, $2p4s$, $2p4p$,} \\
\multicolumn{6}{l}{Ni XXVII: $\lambda_{nl}$ - 1.0, for 1s to 4p } \\
\multicolumn{6}{l}{Ni XXVIII: No correlations; $\lambda_{nl}$ - 1.0, for 1s to 4f } \\
\end{tabular}
\end{table}

\begin{table}
\caption{Radiative decay rates, $A_{ji}$, in sec$^{-1}$ for allowed 
transitions to the ground level, $1s^2~^1S_0$ for Ni XXVII and 
$1s~^2S_{1/2}$ for Ni XXVIII (specified in the column title). 
 }
\scriptsize
\begin{tabular}{llll}
\hline
\multicolumn{1}{c}{Target} & \multicolumn{1}{c}{$A_{fi}$} &
\multicolumn{1}{c}{Target} & \multicolumn{1}{c}{$A_{fi}$} \\
\multicolumn{1}{c}{State} & \multicolumn{1}{c}{($s^{-1}$)} & 
\multicolumn{1}{c}{State} & \multicolumn{1}{c}{($s^{-1}$)} \\
\hline
\multicolumn{4}{c}{Ni XXVII:~GD-$1s^2~^1S_0$} \\
\hline
$1s2p(^3P^o_3)$ & 7.21(13) & $1s3p(^3P^o_3)$ & 2.17(13) \\
$1s2p(^3P^o_1)$ & 6.23(14) & $1s3p(^3P^o_1)$ & 1.80(14) \\
\hline
\multicolumn{4}{c}{Ni~XXVIII:~GD-$1s~^2S^{1/2}$} \\
\hline
$2p(^2P^o_{1/2})$ & 3.79(14) & $3p(^2P^o_{3/2})$ & 9.73(13) \\
$2p(^2P^o_{3/2})$ & 3.80(14) & $4p(^2P^o_{1/2})$ & 3.23(13) \\
$3p(^2P^o_{1/2})$ & 9.34(13) & $4p(^2P^o_{3/2})$ & 3.50(13) \\
\hline
\end{tabular}
\end{table}

\pagebreak

\begin{table}
\caption{Total recombination rate coefficients $\alpha_R(T)$ for Ni~XXVI,
Ni~XXVII and Ni~XXVIII. }
\scriptsize
\begin{tabular}{crrrcrrr}
\hline
$log_{10}T$ & \multicolumn{3}{c}{$\alpha_R(cm^3s^{-1})$} &
$log_{10}T$ & \multicolumn{3}{c}{$\alpha_R(cm^3s^{-1})$}\\
(K) & \multicolumn{1}{c}{Ni XXVI} & \multicolumn{1}{c}{Ni XXVII} &
\multicolumn{1}{c}{Ni XXVIII} & (K) & \multicolumn{1}{c}{Ni XXVI} &
\multicolumn{1}{c}{Ni XXVII} & \multicolumn{1}{c}{Ni XXVIII} \\
\hline
  1.0& 2.86E-08&3.28E-08&3.72E-08 &
  5.1& 1.21E-10&1.50E-10&1.80E-10 \\
  1.1& 2.54E-08&2.92E-08&3.31E-08 &
  5.2& 1.04E-10&1.29E-10&1.56E-10 \\
  1.2& 2.26E-08&2.60E-08&2.95E-08 &
  5.3& 8.92E-11&1.11E-10&1.35E-10 \\
  1.3& 2.01E-08&2.31E-08&2.62E-08 &
  5.4& 7.65E-11&9.61E-11&1.17E-10 \\
  1.4& 1.78E-08&2.05E-08&2.33E-08 &
  5.5& 6.55E-11&8.27E-11&1.01E-10 \\
  1.5& 1.58E-08&1.82E-08&2.07E-08 &
  5.6& 5.59E-11&7.12E-11&8.69E-11 \\
  1.6& 1.40E-08&1.62E-08&1.84E-08 &
  5.7& 4.77E-11&6.11E-11&7.50E-11 \\
  1.7& 1.24E-08&1.43E-08&1.63E-08 &
  5.8& 4.07E-11&5.24E-11&6.46E-11 \\
  1.8& 1.10E-08&1.27E-08&1.44E-08 &
  5.9& 3.46E-11&4.49E-11&5.56E-11 \\
  1.9& 9.74E-09&1.12E-08&1.28E-08 &
  6.0& 2.93E-11&3.84E-11&4.78E-11 \\
  2.0& 8.60E-09&9.94E-09&1.13E-08 &
  6.1& 2.48E-11&3.28E-11&4.10E-11 \\
  2.1& 7.60E-09&8.78E-09&1.00E-08 &
  6.2& 2.10E-11&2.80E-11&3.53E-11 \\
  2.2& 6.70E-09&7.76E-09&8.85E-09 &
  6.3& 1.77E-11&2.39E-11&3.02E-11 \\
  2.3& 5.90E-09&6.84E-09&7.81E-09 &
  6.4& 1.49E-11&2.03E-11&2.58E-11 \\
  2.4& 5.20E-09&6.03E-09&6.89E-09 &
  6.5& 1.25E-11&1.72E-11&2.21E-11 \\
  2.5& 4.58E-09&5.31E-09&6.07E-09 &
  6.6& 1.05E-11&1.46E-11&1.88E-11 \\
  2.6& 4.01E-09&4.67E-09&5.35E-09 &
  6.7& 8.72E-12&1.23E-11&1.60E-11 \\
  2.7& 3.52E-09&4.10E-09&4.70E-09 &
  6.8& 7.26E-12&1.04E-11&1.36E-11 \\
  2.8& 3.09E-09&3.60E-09&4.14E-09 &
  6.9& 6.03E-12&8.75E-12&1.16E-11 \\
  2.9& 2.71E-09&3.16E-09&3.63E-09 &
  7.0& 5.00E-12&7.35E-12&9.78E-12 \\
  3.0& 2.37E-09&2.77E-09&3.19E-09 &
  7.1& 4.17E-12&6.18E-12&8.25E-12 \\
  3.1& 2.07E-09&2.43E-09&2.80E-09 &
  7.2& 3.51E-12&5.22E-12&6.97E-12 \\
  3.2& 1.81E-09&2.13E-09&2.45E-09 &
  7.3& 3.01E-12&4.44E-12&5.85E-12 \\
  3.3& 1.58E-09&1.86E-09&2.15E-09 &
  7.4& 2.63E-12&3.82E-12&4.90E-12 \\
  3.4& 1.38E-09&1.63E-09&1.88E-09 &
  7.5& 2.32E-12&3.30E-12&4.10E-12 \\
  3.5& 1.21E-09&1.42E-09&1.65E-09 &
  7.6& 2.05E-12&2.86E-12&3.41E-12 \\
  3.6& 1.05E-09&1.24E-09&1.44E-09 &
  7.7& 1.78E-12&2.47E-12&2.83E-12 \\
  3.7& 9.14E-10&1.08E-09&1.26E-09 &
  7.8& 1.53E-12&2.10E-12&2.34E-12 \\
  3.8& 7.95E-10&9.44E-10&1.10E-09 &
  7.9& 1.28E-12&1.77E-12&1.92E-12 \\
  3.9& 6.91E-10&8.22E-10&9.59E-10 &
  8.0& 1.06E-12&1.46E-12&1.58E-12 \\
  4.0& 6.00E-10&7.16E-10&8.37E-10 &
  8.1& 8.52E-13&1.19E-12&1.28E-12 \\
  4.1& 5.21E-10&6.23E-10&7.29E-10 &
  8.2& 6.76E-13&9.60E-13&1.04E-12 \\
  4.2& 4.52E-10&5.42E-10&6.37E-10 &
  8.3& 5.29E-13&7.64E-13&8.36E-13 \\
  4.3& 3.92E-10&4.71E-10&5.54E-10 &
  8.4& 4.09E-13&6.01E-13&6.70E-13 \\
  4.4& 3.39E-10&4.09E-10&4.82E-10 &
  8.5& 3.13E-13&4.69E-13&5.33E-13 \\
  4.5& 2.94E-10&3.55E-10&4.20E-10 &
  8.6& 2.37E-13&3.64E-13&4.22E-13 \\
  4.6& 2.54E-10&3.08E-10&3.65E-10 &
  8.7& 1.79E-13&2.80E-13&3.32E-13 \\
  4.7& 2.19E-10&2.67E-10&3.17E-10 &
  8.8& 1.34E-13&2.14E-13&2.59E-13 \\
  4.8& 1.89E-10&2.32E-10&2.76E-10 &
  8.9& 1.00E-13&1.63E-13&2.02E-13 \\
  4.9& 1.63E-10&2.00E-10&2.39E-10 &
  9.0& 7.44E-14&1.24E-13&1.56E-13 \\
  5.0& 1.41E-10&1.73E-10&2.08E-10 &
  & & & \\
\hline
\end{tabular}
\end{table}


\clearpage

%
%

\def\amp{{Adv. At. Molec. Phys.}\ }
\def\apj{{ Astrophys. J.}\ }
\def\apjs{{Astrophys. J. Suppl. Ser.}\ }
\def\apjl{{Astrophys. J. (Letters)}\ }
\def\aj{{Astron. J.}\ }
\def\aa{{Astron. Astrophys.}\ }
\def\aasup{{Astron. Astrophys. Suppl.}\ }
\def\adndt{{At. Data Nucl. Data Tables}\ }
\def\cpc{{Comput. Phys. Commun.}\ }
\def\jqsrt{{J. Quant. Spect. Radiat. Transfer}\ }
\def\jpb{{Journal Of Physics B}\ }
\def\pasp{{Pub. Astron. Soc. Pacific}\ }
\def\mn{{Mon. Not. R. astr. Soc.}\ }
\def\pra{{Physical Review A}\ }
\def\prl{{Physical Review Letters}\ }
\def\zpds{{Z. Phys. D Suppl.}\ }
\def\adndt{Atomic Data And Nuclear Data Tables}

%

\begin{figure}
\epsscale{.80}
\plotone{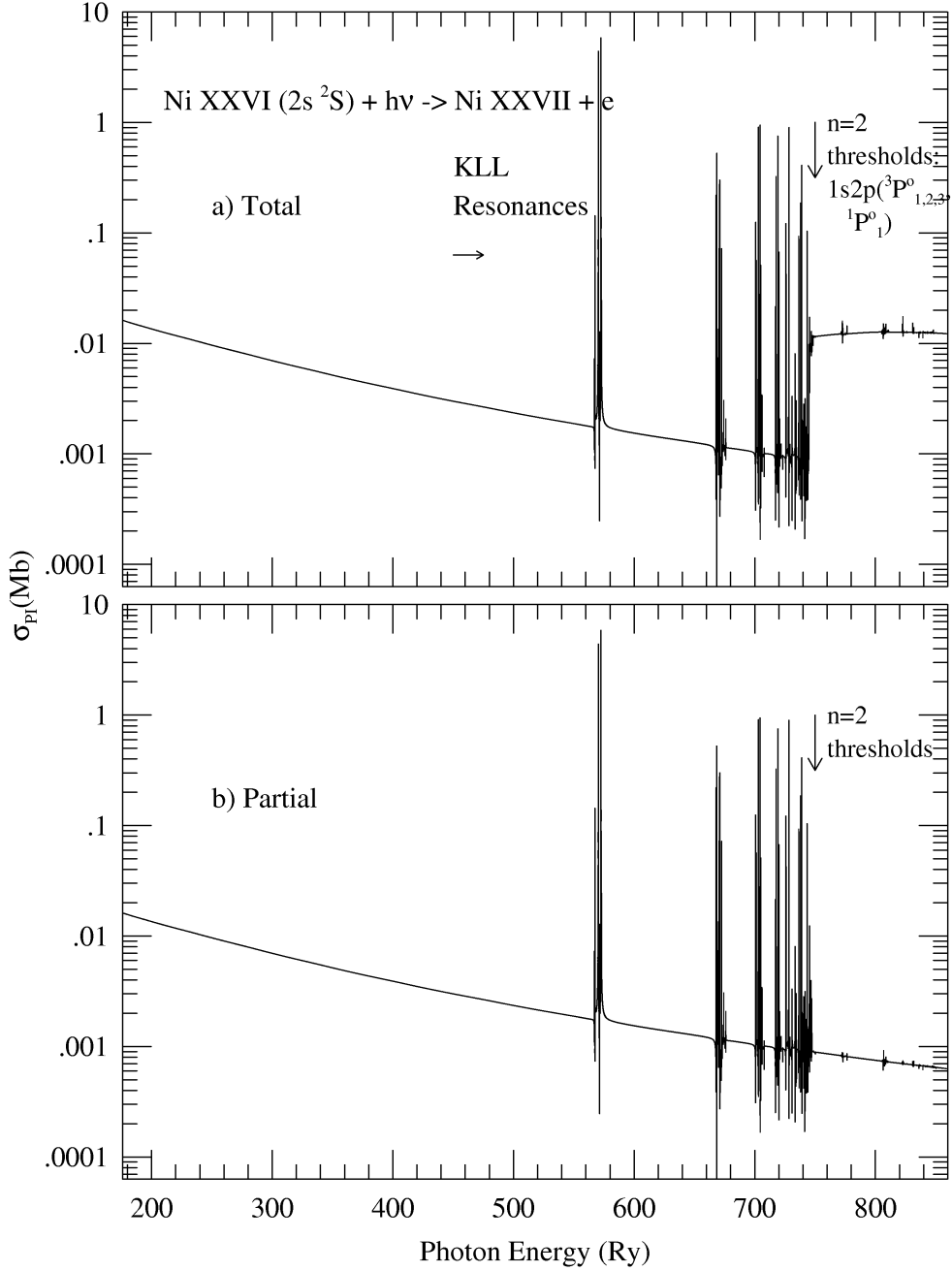}
\caption{Photoionization cross sections ($\sigma_{PI}$) of the ground 
level $1s^2 \ 2s \ (^2S_{1/2})$ of Ni~XXVI: (a) Total cross section; 
the large jump around n = 2 thresholds ($\sim$ 735 Ry) is the K-shell 
ionization edge (h$\nu$ + 1s$^2$2$l$ $\rightarrow$ 1s2$l$ + e). (b) Partial 
cross section into the ground level $1s^2 \ (^1S_0)$ 
of Ni~XXVII; note that the jump is no longer present and the cross 
section is continuous across the n = 2 thresholds of Ni~XXVII.}
\end{figure}

\begin{figure}
\epsscale{.80}
\plotone{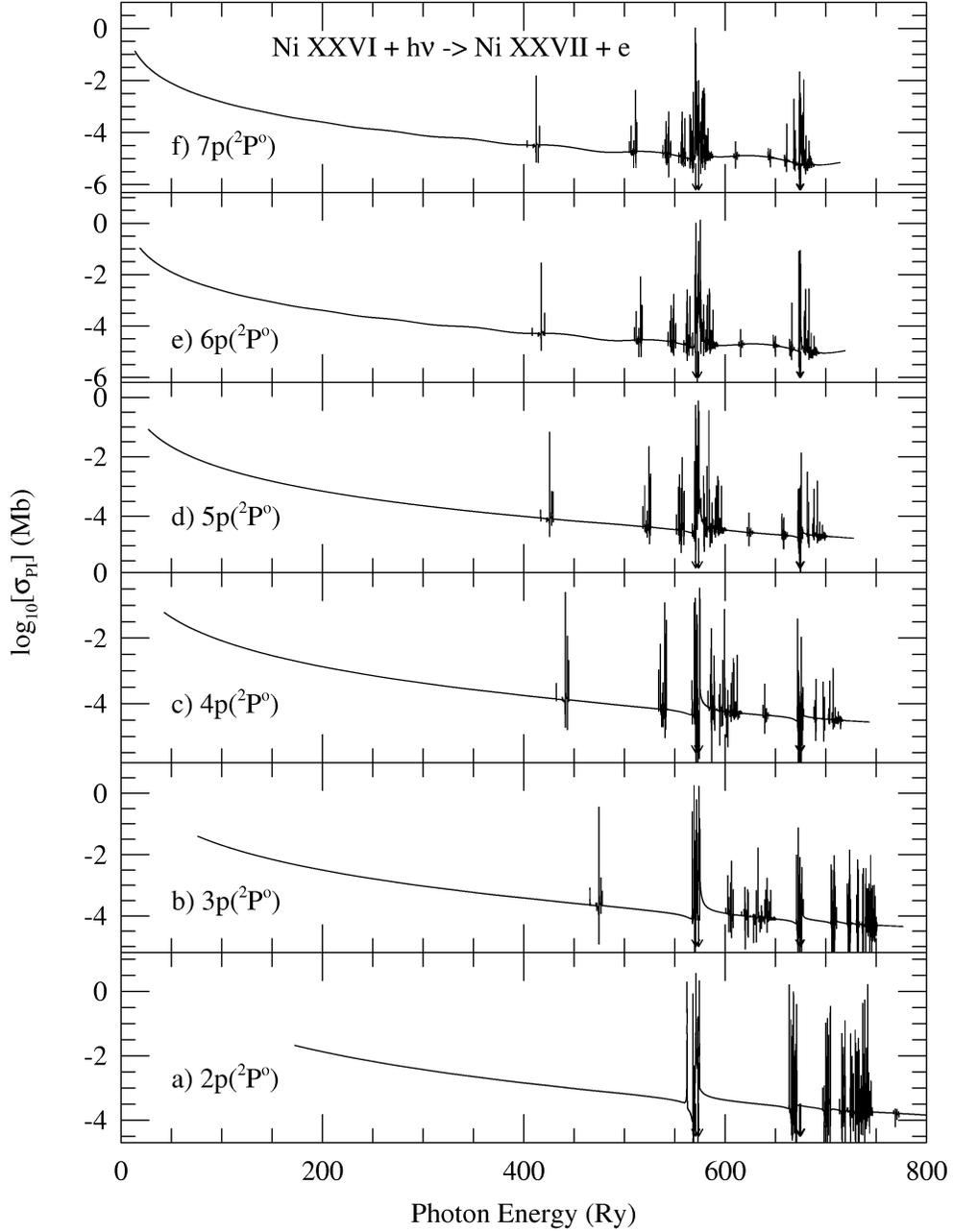}
\caption{Partial photoionization cross sections of the Rydberg series 
of levels, $1s^2np(^2P^o_{1/2})$ with 2 $\leq n \leq$ 7, of Ni~XXVI into 
the ground level $1s^2(^1S_0)$ of Ni~XXVII. Prominent PEC 
({\it photoexcitation-of-core}) resonances are seen (pointed by arrows) 
at about 571 Ry of excited core levels $1s2p~^3P^o_1,^1P^o_1$ and about 
674 Ry of levels $1s3p~^3P^o_1,^1P^o_1$.}
\end{figure}

\begin{figure}
\epsscale{.80}
\plotone{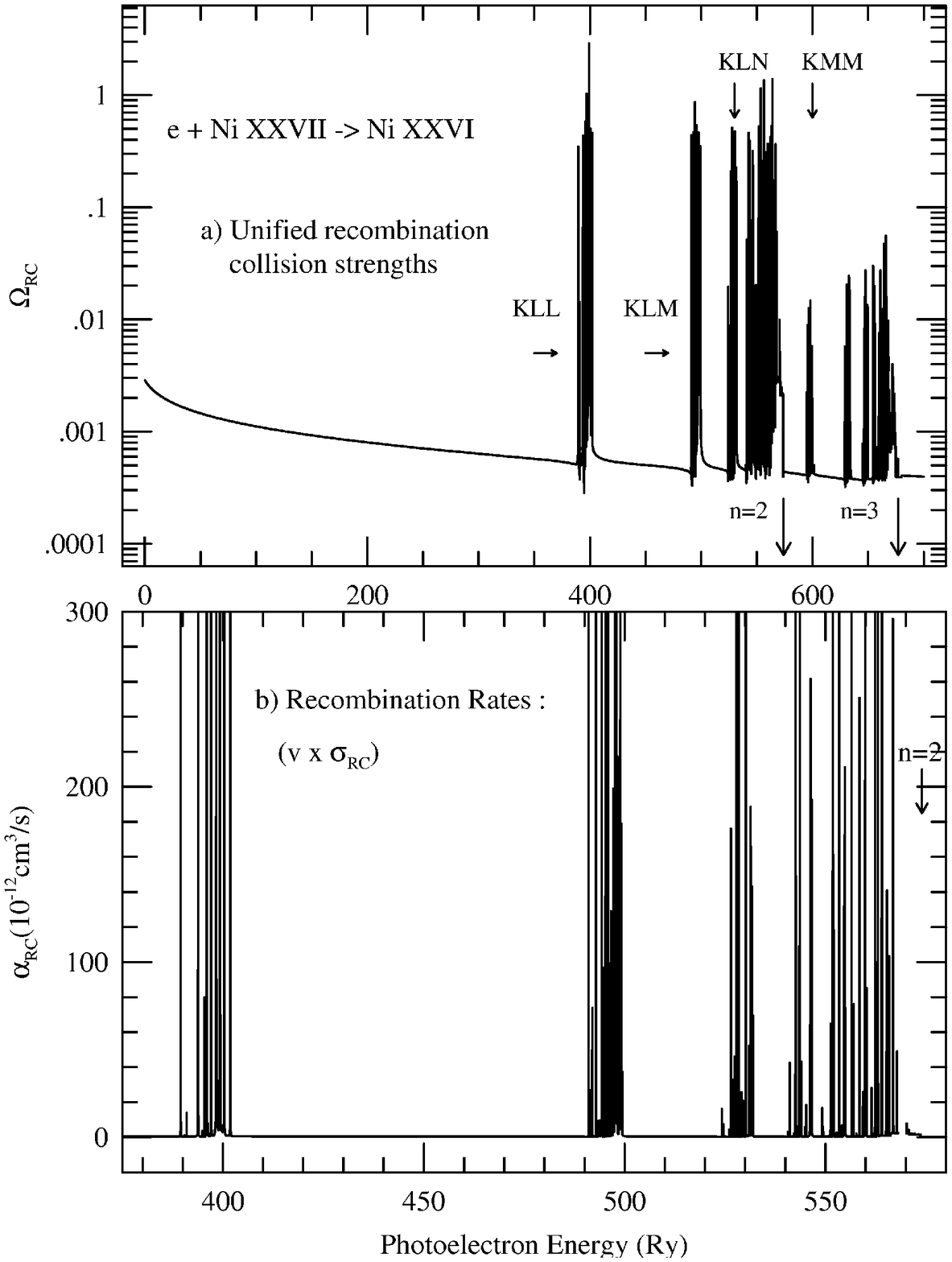}
\caption{(a) Total unified (e + ion) recombination collision strengths,
$\Omega_{RC}$ and (b) unified recombination rate coefficients, 
$\alpha_{RC}(E)$, with photoelectron energy of Ni XXVI. Note 
separated resonance complexes, KLL, KLM, etc of n = 2 and KMM, KMN etc. 
of n = 3 thresholds. $\alpha_{RC}(E)$, convolved with a bandwidth is a
measurable quantity.} 
\end{figure}

\begin{figure}
\epsscale{.80}
\plotone{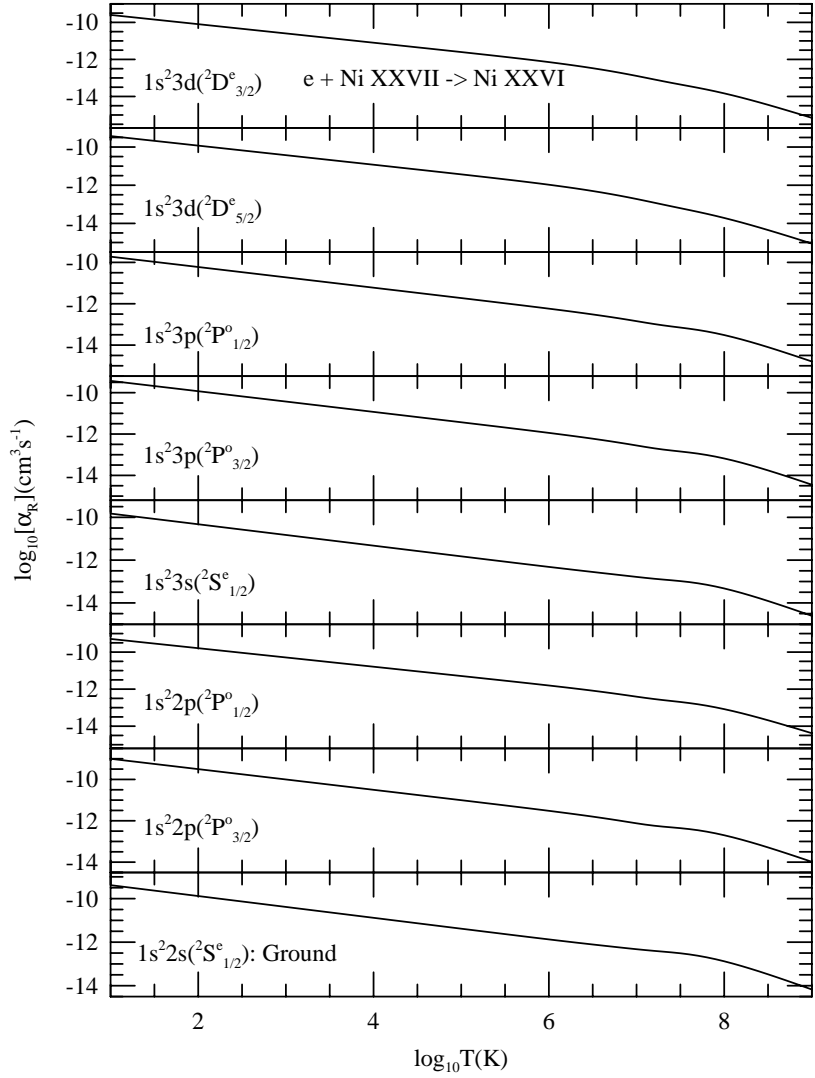}
\caption{ Level-specific recombination rate coefficients for Ni~XXVI
recombining to ground and excited n=2, 3 levels.}
\end{figure}

\begin{figure}
\epsscale{.80}
\plotone{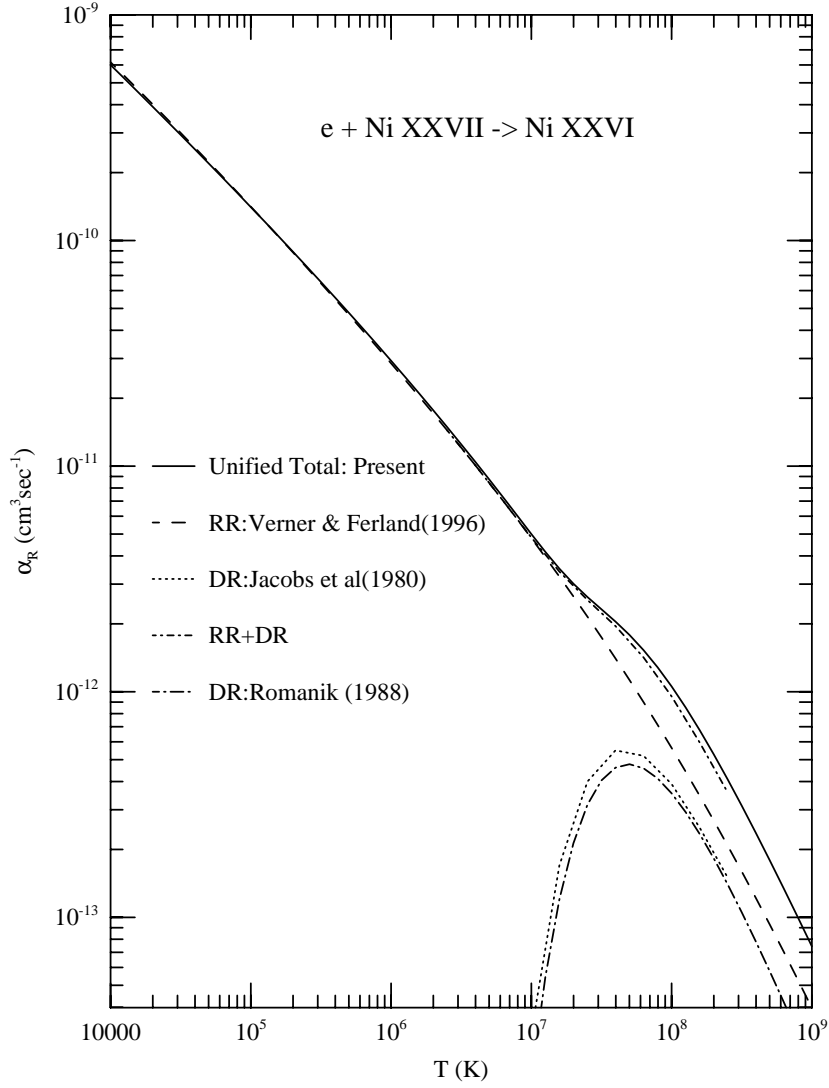}
\caption{Total unified recombination rate coefficients, $\alpha_R(T)$, 
for Ni~XXVI: total unified (solid curve), RR rates (dash) by Verner and 
Ferland (1996), DR rates (dotted) by Jacobs et al (1980) and DR rates
(chain dash) by Romanik (1988), and sum of RR+DR (dot-dash). }
\end{figure}

\begin{figure}
\epsscale{.80}
\plotone{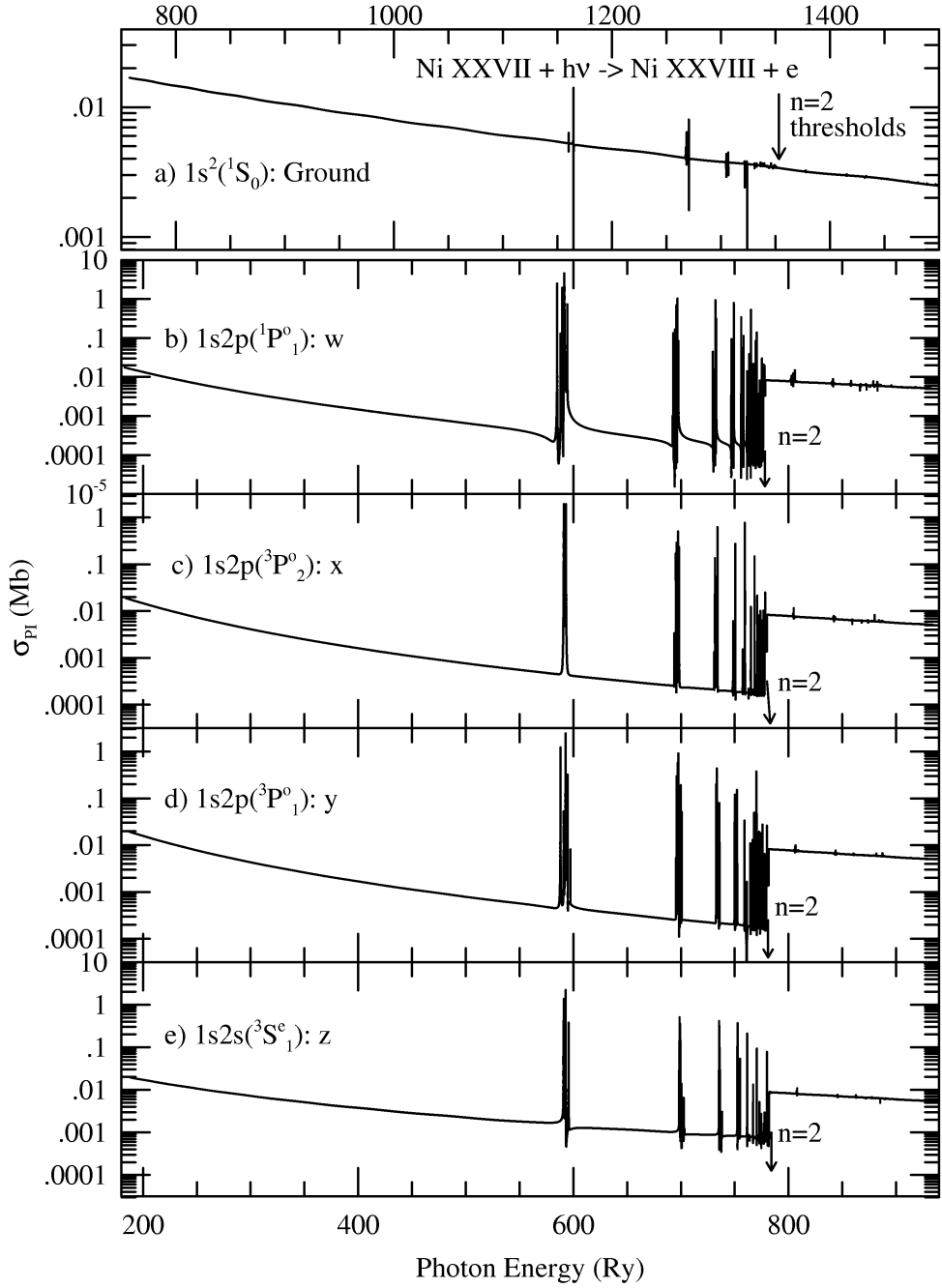}
\caption{Level-specific photoionization cross sections of (a) the 
ground $1s^2 \ (^1S_0)$, and excited (b) 1s2p($^1P^o_1$)-w, (c)
1s2p($^3P^o_2$)-x, (d) 1s2p($^3P^o_1$)-y, (e) 1s2s($^3S_1$)-z levels of 
Ni~XXVII. The excited levels correspond to the prominent X-ray lines: 
resonance (w), intercombination (y), and forbidden(x,z) and show K-shell
ionization edge at n = 2 thresholds.}
\end{figure}

\begin{figure}
\epsscale{.80}
\plotone{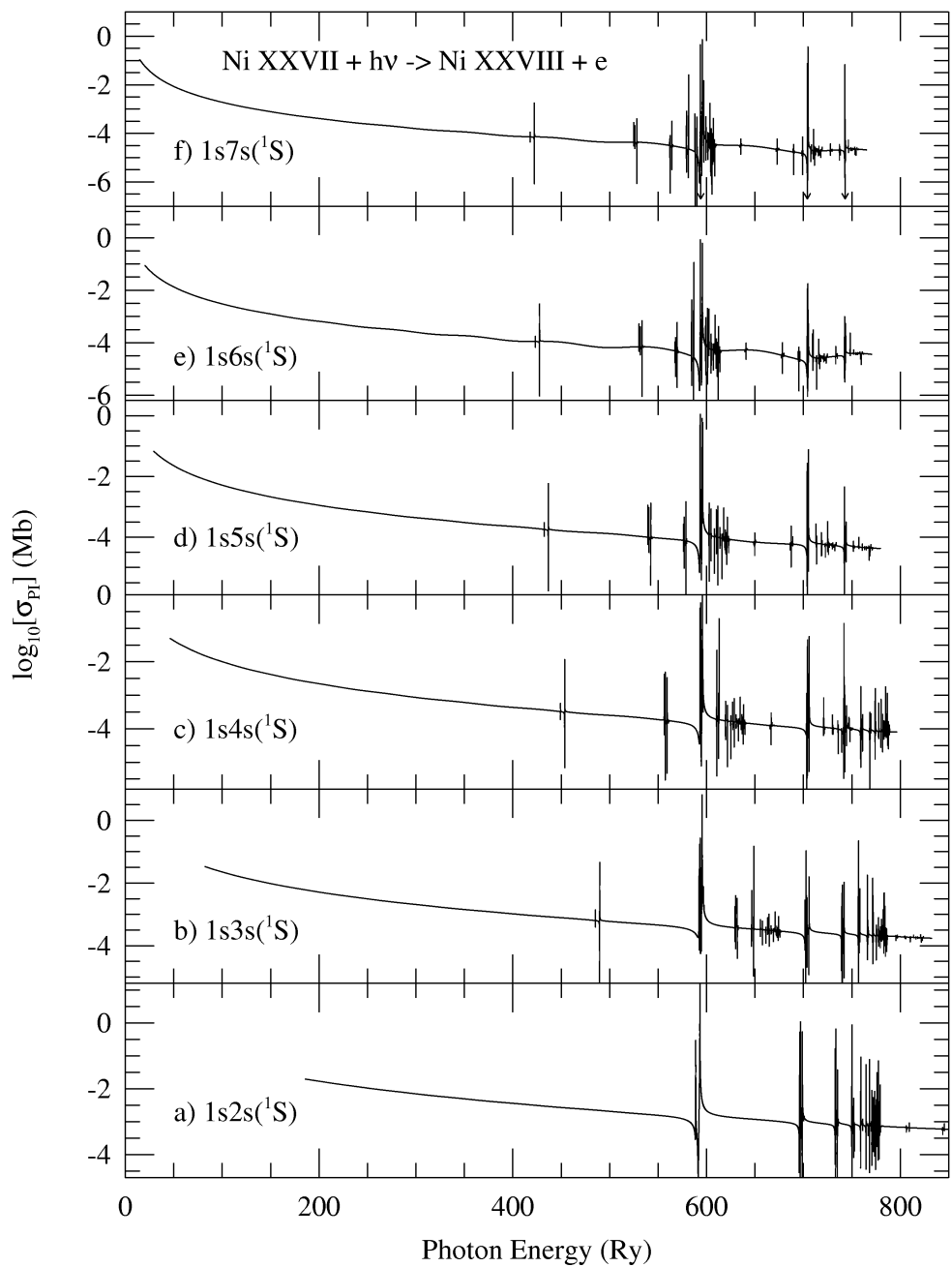}
\caption{Partial photoionization cross sections of the Rydberg series
of levels $1sns(^1S_{0})$, 2 $\leq n \leq$ 7, of Ni~XXVII into the 
ground level $1s(^2S_{1/2})$ of the core, displaying PEC 
({\it photoexcitation-of-core}) resonances at about 594, 704, 743 Ry 
(marked by arrows in the top panel) of core levels $^2P^o_{1/2,3/2}$ of 
configurations $2p$, $3P$, and $4p$.}
\end{figure}

\begin{figure}
\epsscale{.80}
\plotone{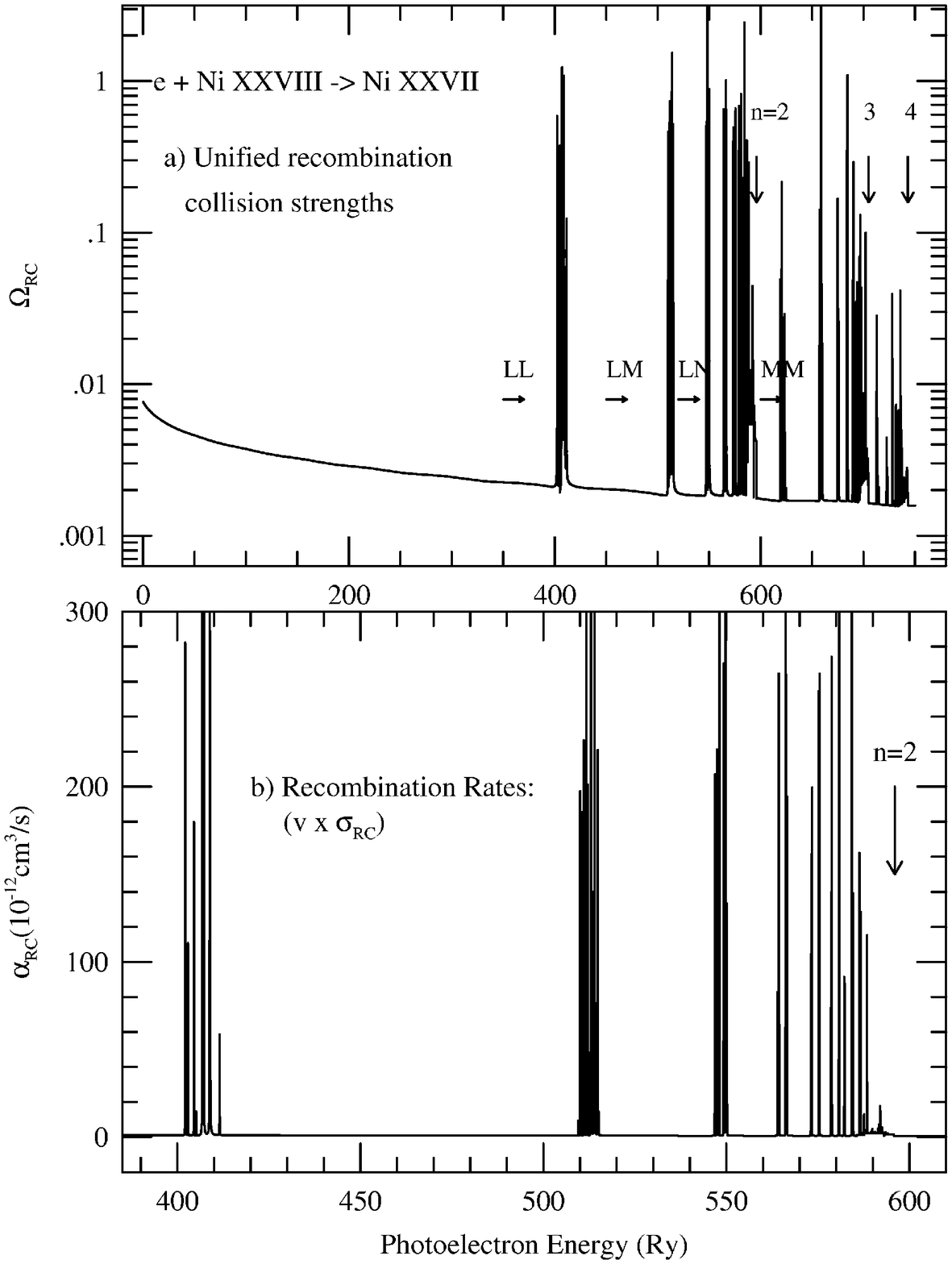}
\caption{(a) Total unified (e + ion) recombination collision strengths,
$\Omega_{RC}$ and (b) unified recombination rate coefficients, 
$\alpha_{RC}(E)$ with photoelectron energy of Ni XXVII. Note the
separated resonance complexes, LL, LM, etc of n = 2 and MM, MN etc. 
of n = 3 and NN, NO etc of n = 4 thresholds. $\alpha_{RC}(E)$, convolved 
with a bandwidth, is a measurable quantity.} 
\end{figure}

\begin{figure}
\epsscale{.80}
\plotone{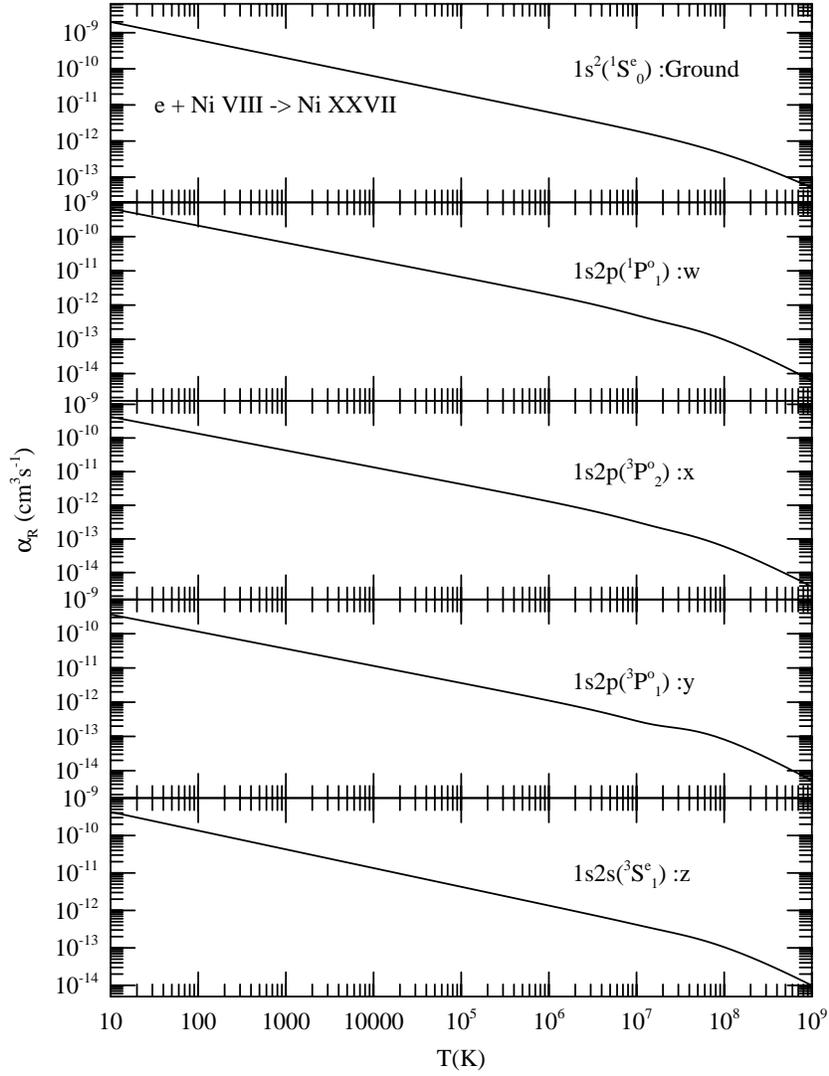}
\caption{ Level-specific recombination rate coefficients for Ni XXVII 
into the ground and excited n = 2 levels responsible for the prominent 
X-ray w, x, y, and z lines.}
\end{figure}

\begin{figure}
\epsscale{.80}
\plotone{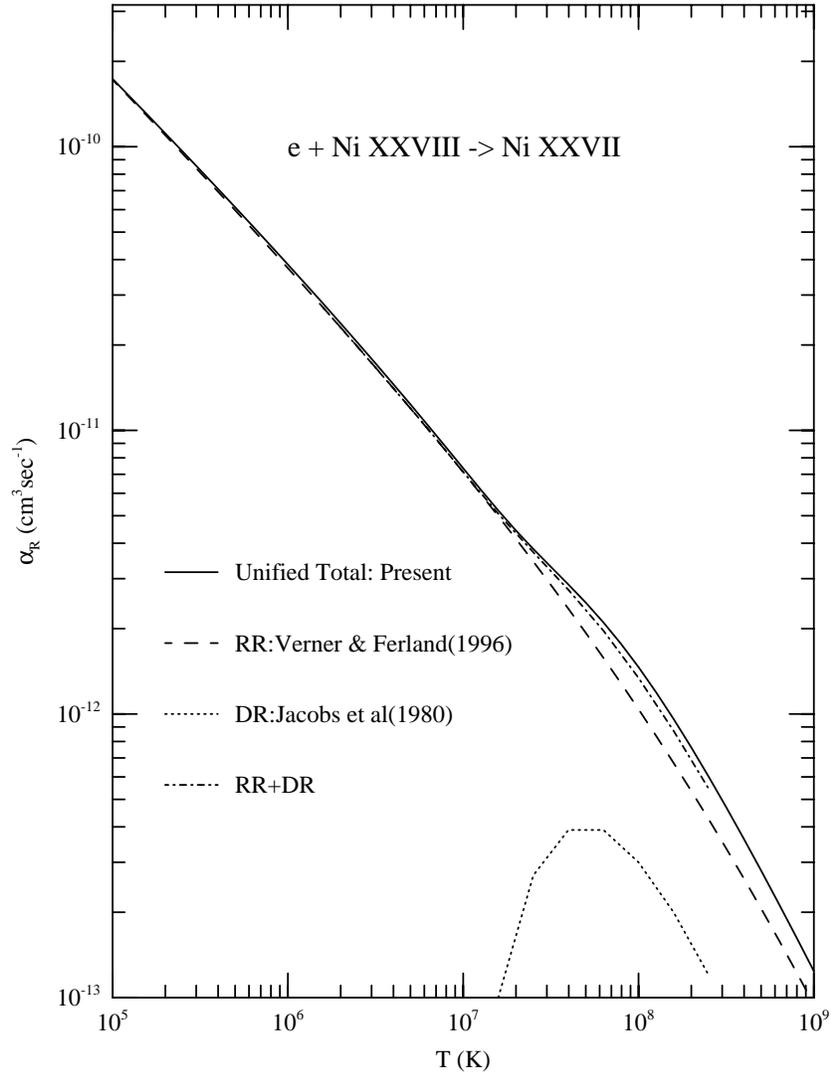}
\caption{Total unified recombination rate coefficients, $\alpha_R(T)$,
for Ni~XXVII: present unified total (solid), RR rates (dash) by Verner 
and Ferland (1996), DR rates (dot) by Jacobs et al. (1980).
}
\end{figure}

\end{document}